# Uncovering many-body correlations in nanoscale nuclear spin baths by central spin decoherence

Wen-Long Ma[1,2,3,†], Gary Wolfowicz[4,5,†], Nan Zhao[6], Shu-Shen Li[1,3], John J. L. Morton[4,7*] & Ren-Bao Liu[2,8,9*]

1. State Key Laboratory of Superlattices and Microstructures, Institute of Semiconductors, Chinese Academy of Sciences, Beijing, China
2. Department of Physics, The Chinese University of Hong Kong, Shatin, N. T., Hong Kong, China
3. Synergetic Innovation Center of Quantum Information & Quantum Physics, University of Science and Technology of China, Hefei, Anhui 230026, China
4. London Centre for Nanotechnology, University College London, London WC1H 0AH, United Kingdom
5. Department of Materials, Oxford University, Oxford OX1 3PH, United Kingdom
6. Beijing Computational Science Research Center, Beijing, China
7. Department of Electronic & Electrical Engineering, University College London, London WC1E 7JE, United Kingdom
8. Center for Quantum Coherence, The Chinese University of Hong Kong, Shatin, N.T., Hong Kong, China
9. Institute of Theoretical Physics, The Chinese University of Hong Kong, Shatin, N.T., Hong Kong, China

† These authors contributed equally to this work.
* Corresponding authors.

**ABSTRACT:**

**Many-body correlations can yield key insights into the nature of interacting systems; however, detecting them is often very challenging in many-particle physics, especially in nanoscale systems. Here, taking a phosphorus donor electron spin in a natural-abundance $^{29}$Si nuclear spin bath as our model system, we discover both theoretically and experimentally that many-body correlations in nanoscale nuclear spin baths produce identifiable signatures in the decoherence of the central spin under multiple-pulse dynamical decoupling control. We find that when the number of decoupling $\pi$-pulses is odd, central spin decoherence is primarily driven by second-order nuclear spin correlations (pairwise flip-flop processes). In contrast, when the number of $\pi$-pulses is even, fourth-order nuclear spin correlations (diagonal interaction renormalized pairwise flip-flop processes) are principally responsible for the central spin decoherence. Many-body correlations of different orders can thus be selectively detected by central spin decoherence under different dynamical decoupling controls, providing a useful approach to probing many-body processes in nanoscale nuclear spin baths.**

**FULL TEXT:**

Decoherence of a central spin in a solid-state environment is not only an ideal model problem for understanding the foundation of quantum physics [1-3] but also a critical issue in a number of quantum technologies including spin-based quantum information processing [4, 5] and ultrasensitive magnetometry [6-10]. For example, decoherence from the environmental spin bath is often a limiting factor when using systems such as phosphorous donors in silicon [11-16], semiconductor quantum dots [17, 18] and nitrogen-vacancy centers in diamond [19, 20], as quantum bits or sensors. Studying central spin decoherence caused by environmental fluctuations or elementary excitations may yield key insights into the nature of many-body interactions in the environment. Furthermore, dynamical control over the central spin can affect the dynamics of the environment in a detectable manner [8, 18]. In the light of these ideas, exploiting central spin decoherence for sensing single nuclear spins or nuclear spin clusters in spin baths has been theoretically proposed [6-8] and experimentally demonstrated [9,10]. Recently, this idea has been pushed to new depths: theoretical studies show that the central spin decoherence can be a novel probe to many-body physics, in particular, phase transitions in spin baths [21-24]. Multiple-spin correlations are one of the essential characteristics in spin baths [11-20], but detection of such correlations is a long-standing challenge in many-body physics. Here we address this problem with the first experimental demonstration of detection of many-body correlations via central spin decoherence, laying a foundation for studying many-body physics and phase transitions in spin baths [21-24].

Previous approaches to studying multiple-particle correlations include the use of nonlinear optical spectroscopy of excitons in semiconductors [25-28], nuclear magnetic resonance (NMR) spectroscopy of nuclear spins in molecules [29], and the generalisation of multi-dimensional

NMR to optical spectroscopy [30, 31]. Nevertheless, the detection and characterization of many-body correlations in nanoscale systems [32, 33] remain highly challenging due to the weak signals in such small systems. In this article, we find that many-body correlations in nanoscale nuclear spin baths have identifiable effects on the decoherence of a central spin. This enables us to propose and implement a scheme to detect many-body correlations of different orders in the nuclear spin bath through monitoring the central spin decoherence. We can distinguish the second-order nuclear spin correlations from the fourth-order nuclear spin correlations by applying different numbers of pulses in dynamical decoupling control of the central spin. Our proposal is particularly suited for the detection of many-body correlations in nanoscale systems.

## Results

**System and model.** We consider the electron spin ($S = 1/2$) of a phosphorus donor localized in silicon as the central spin (Fig. 1a). This donor electron spin is coupled with a $^{29}$Si nuclear spin bath ($I = 1/2$ and natural abundance of 4.7% throughout the host lattice) by the contact hyperfine interactions and dipolar interactions [14]. In a strong external magnetic field the Zeeman energies of the donor spin and nuclear spins are conserved, so the total Hamiltonian can be written in the secular form [12, 13]

$$H = \omega_e S_z + S_z \sum_i A_i I_i^z - \omega_n \sum_i I_i^z + \sum_{i<j} D_{ij} (I_i^+ I_j^- + I_i^- I_j^+ - 4 I_i^z I_j^z), \tag{1}$$

where $\omega_{e/n} = \gamma_{e/n} B$ is the Larmor frequency of the donor electron spin /bath nuclear spins, $\gamma_{e/n}$ is the gyromagnetic ratio of the donor electron spin /bath nuclear spins, and $B$ is the external magnetic field applied along the $z$-axis. The coupling coefficient between the donor spin and the $i$-th nuclear spin is $A_i = \gamma_e \gamma_n [8\pi/3 |\psi(\mathbf{R}_i)|^2 + \theta(|\mathbf{R}_i| - r_0)(3\cos^2\theta_i - 1)/|\mathbf{R}_i|^3]$, where $\psi(\mathbf{R}_i)$ is the donor electron wave function at the position of the $i$-th nuclear spin, $\theta(r)$ is the Heaviside

step function and $\theta_i$ is the angle between the nuclear spin position vector $\mathbf{R}_i$ and the magnetic field vector $\mathbf{B}$. In this expression of $A_i$, the first part represents the contact hyperfine interaction while the second part represents the dipolar interaction which starts contributing for $|\mathbf{R}_i| > r_0 = 2$ nm. The dipolar interaction between the nuclear spins is $D_{ij} = \gamma_n^2 (3\cos^2 \theta_{ij} - 1)/4|\mathbf{R}_{ij}|^3$, where $\theta_{ij}$ is the angle between $\mathbf{R}_{ij} = \mathbf{R}_i - \mathbf{R}_j$ and $\mathbf{B}$.

We assume that the donor electron spin is initially prepared in the coherent state $(|+\rangle + |-\rangle)/\sqrt{2}$ by a $\pi/2$-rotation (with +/− being spin-up/down along the magnetic field direction). In the subsequent evolution, the central spin suffers decoherence as a result of its coupling to the nuclear spin bath. However, by applying dynamical decoupling (DD) control [34, 35] to the central spin (consisting of a sequence of $\pi$-flips at times $\{t_1, t_2 \cdots t_n\}$), we can reduce its sensitivity to the bath in general while selectively enhancing the effect of certain multiple-spin dynamics [8]. With DD, the restored central spin coherence following a total evolution time $T$ is

$$L_{+,-}^n(T) = \sum_J P_J \langle J | (U_+^n(T))^\dagger U_-^n(T) | J \rangle, \qquad (2)$$

with

$$U_\pm^n(T) = e^{-i[V + (-1)^n H_0](T - t_n)} \cdots e^{-i(V \mp H_0)(t_2 - t_1)} e^{-i(V \pm H_0) t_1}, \qquad (3)$$

where $H_0 = \frac{1}{2} \sum_i A_i I_i^z$ and $V = \sum_{i<j} D_{ij}(I_i^+ I_j^- + I_i^- I_j^+ - 4 I_i^z I_j^z)$. Here, the nuclear Zeeman term $\omega_n \sum_i I_i^z$ is dropped since it has no contribution to the spin decoherence. The nuclear spin bath is assumed to be in an infinite-temperature (fully mixed) state with density matrix $\rho_0 = \sum_J |J\rangle\langle J|/2^M$ where $|J\rangle$ is an eigenstate of $\sum_i I_i^z$ and $M$ being the number of nuclear spins in the bath.

We consider two families of DD sequences: Carr-Purcell-Meiboom-Gill (CPMG) [36-38] and Uhrig dynamical decoupling (UDD) [39, 40] (Fig. 2a). An *n*-pulse CPMG sequence periodically flips the central spin at time $t_c = (2c-1)T/2n$, while *n*-pulse UDD flips the central spin at time $t_c = T\sin^2[c\pi/(2n+2)]$, where $T$ is the total evolution time and $c = 1\cdots n$. It should be noted that CPMG and UDD are equivalent for $n \leq 2$, and for $n = 1$ simply correspond to the Hahn echo.

**Many-body correlation effects on central spin decoherence.** According to the linked-cluster expansion (LCE) theorem in many-body physics [41], the quantum evolution of a nuclear spin bath can be factorized into contributions of different orders of irreducible many-body correlations, namely,

$$L^n_{+,-}(T) = \exp(\langle V_1 \rangle + \langle V_2 \rangle + \langle V_3 \rangle + \langle V_4 \rangle + \cdots), \quad (4)$$

with the *l*-th order many-body correlation

$$\langle V_l \rangle = \frac{1}{l!} \int_C dt_1 \cdots \int_C dt_k \langle J | \hat{T}_C \{V(t_1) \cdots V(t_l)\} | J \rangle, \quad (5)$$

where $\hat{T}_C$ is the time-ordering operator along the contour $C (0 \to T \to 0)$, and $V(t) = \exp(iH_0 t) V \exp(-iH_0 t)$ is the intra-bath coupling in the interaction picture. We show some examples of the expansion terms diagrammatically in Fig. 1b (see Fig. S1 in Supplementary Information for more diagrams). Here we assume the nuclear spin bath starts from a pure product state $|J\rangle$. The thermal ensemble results can be obtained by sampling over different initial states and then taking a statistical average.

For each LCE term, the real part contributes to the spin decoherence while the imaginary part just produces a coherent phase shift (corresponding to self-energy renormalization of the probe spin). Under CPMG-$n$ or UDD-$n$ control, the first-order LCE term ($l = 1$) vanishes due to the contour integral. The second-order LCE term ($l = 2$) corresponds to the pairwise flip-flop processes in the nuclear spin bath, in which the bath dynamics is approximated as a product of evolutions of nuclear spin pairs [15, 17, 18]. Previous studies identified this term as the main cause of spin decoherence for the free-induction decay and Hahn echo in the strong magnetic field regime [15, 17, 18]. The pairwise flip-flop processes of nuclear spins $i, j$ can be mapped to the precession of a pseudospin $\sigma_{ij}$ about a pseudofield $h_{ij}^{\pm} = (D_{ij}, 0, \omega_{ij}/2)$ conditioned on the central spin state $|\pm\rangle$ [17] (see Supplementary Information), where $\omega_{ij} = (A_i - A_j)/2$ is the energy cost of the flip-flop process. If the central spin is under CPMG-$n$ control, we have $\text{Re}\langle V_2^{\text{odd}}\rangle = \sum_{ij} 4 D_{ij}^2 \omega_{ij}^{-2} \left[ 4\cos(\omega_{ij}t) - \cos(2\omega_{ij}t) - 3 \right]$ when $n$ is odd, but $\text{Re}\langle V_2^{\text{even}}\rangle = 0$ when $n$ is even (see the schematics in Fig. 3a), where $t = T/2n$. For UDD-$n$ control, the real part of second-order LCE term also vanishes when $n$ is even and is nonzero when $n$ is odd (see Supplementary Information for detailed derivations).

For higher-order LCE terms, there are three groups of diagrams: ring diagrams, diagonal-interaction renormalized diagrams, and locked diagrams [41]. Generally, the leading terms of the $l$-th order diagrams are proportional to $(D_{ij}/\omega_{ij})^l$. Due to the random distribution of nuclear spins, the contributions from different nuclear spin clusters add destructively when $l$ is odd but add constructively when $l$ is even. Hence, the odd-order LCE terms contribute negligibly to the spin decoherence.

The central spin decoherence problem can be exactly solved by the cluster-correlation expansion (CCE) method [42]. To identify the contributions of different many-body correlations to the central spin decoherence, we compare the approximate results obtained by the LCE to the exact numerical results obtained by the CCE (Fig. 2b). We see that the second-order pairwise flip-flop LCE term ($V_2$) almost fully reproduces the CCE results for DD controls of odd pulse number, while the contribution of the fourth-order diagonal-interaction renormalized LCE term ($V_{4z}$) coincides with the CCE results for DD controls of even pulse number. This indicates that we can selectively detect either the second-order or fourth-order many-body correlations by choosing an appropriate number of DD control pulses. Similar pulse-number parity effects were theoretically noticed before [38], however, without analyzing the underlying microscopic processes.

The different correlations actually present different central spin decoherence features. In particular, the $V_2$ correlation causes decoherence with a faster initial decay but a longer decay tail ($-\ln|L_{+,-}^{\text{odd}}(T)| \sim T^2$); while the decoherence induced by the $V_{4z}$ correlation is better preserved in the short time regime but decays faster in the long time regime ($-\ln|L_{+,-}^{\text{even}}(T)| \sim T^4$).

It should be pointed out that the LCE-$V_{4z}$ term contains two-body, three-body and four-body nuclear spin correlations (Fig. 1a). The two-body fourth-order correlations have no contribution to decoherence, because the pairwise flip-flop of two nuclear spins is independent of the diagonal interaction between them. The nuclear spin clusters contributing the most to central spin decoherence are those four-spin or three-spin clusters with small inter-nuclei distances (<1 nm), so that the energy cost of the pairwise flip-flop processes of two nuclear spins is significantly changed by the other nuclear spins in the cluster (see Supplementary Information). In the calculations, we consider a bath volume with radius 8 nm from the central spin, corresponding to 5000 nuclear spins. Statistical studies (Fig. 3b) show that there are about $1.8 \times 10^4$ such four-spin clusters and $2.6 \times 10^4$ three-spin clusters in the bath. In Fig. 3c we compare the contributions of different many-body correlations and find that the four-body correlations are the main contribution to the central spin decoherence under DD control of even number of pulses. The three-body correlations are non-zero but relatively small.

**Experimental results.** We have observed the pulse-number parity effect in DD experiments on P-donors in natural Si (Fig. 4). The measured decoherence decays fit well in stretched exponential functions $\exp\left[-T/\tau_{ID}-(T/\tau_{SD})^\lambda\right]$ (see Fig. S3 in Supplementary Information). Here the first term $e^{-T/\tau_{ID}}$ represents the instantaneous diffusion caused by dipolar coupling to other P-donor electron spins in the sample ([P] = 3x10$^{14}$/cm$^3$), and the second term $e^{-(T/\tau_{SD})^\lambda}$ represents the central spin decoherence (spectral diffusion) caused by the $^{29}$Si nuclear spin bath.

In Figs. 4a-b, we show the measured decays, corrected to exclude the instantaneous diffusion (with $\tau_{ID}=10$ ms determined by the initial exponential decay of the raw experimental data in Fig. S3 of Supplementary Information). The measured and calculated results agree well for both CPMG-*n* and UDD-*n* controls, without any adjustable parameters in the calculations. In Figs. 4c-d, we compare the central spin coherence decay time $\tau_{SD}$ and exponent stretching factor $\lambda$ of the measured and numerical data as functions of the pulse number *n*. The quantitative and qualitative agreement is remarkable, the only exception being that the measured decay time $\tau_{SD}$ oscillates with *n* somewhat less strongly than expected. As predicted, the stretching factor $\lambda$ oscillates between about 2 and 4 as *n* increases, meaning that either the second-order correlations or fourth-order correlations contribute dominantly to central spin decoherence. The slight decrease of the stretched exponent $\lambda$ with *n* can be ascribed to the emergence of the "Markovian" decoherence when the coherence time is prolonged to exceed the pairwise flip-flop time and the higher-order many-body correlations become more important [42].

## Discussion

The different signatures of the many-body correlations under DD control of the central spin, in particular the pulse-number parity effect in the number of DD control pulses, provide a useful approach to studying many-body physics in the nuclear spin bath. Note that the parity effect is not affected by the type of DD sequences adopted in this paper- it exists in both CPMG and UDD controls. It is remarkable that the many-body correlations between nuclear spins have sizable effects even at temperatures (a few Kelvin in our experiments) much higher than the coupling strengths between the nuclear spins (a few nano-Kelvin).

The pulse-number parity effect should be observable in a broad range of central spin systems as long as the following conditions are satisfied: (i) *pure dephasing condition*- the external magnetic field should be large so that the energy-non-conserving processes (such as single nuclear spin rotations) are highly suppressed (i.e., the total Hamiltonian can be written in the secular form); (ii) *slow/non-Markovian bath condition* - the couplings between nuclear spins should be much weaker than the inverse decoherence time (under this condition that the LCE terms converge rapidly with increasing orders and the central spin decoherence is mainly induced by the lowest-order non-zero LCE terms).

The detection of many-body correlations may find applications in identifying the structures of molecules. In particular, the pulse-number parity effect can be adopted to tell whether the molecules that form the nuclear spin bath have two-body or higher-order interactions among the nuclei. It should be noted that the current scheme can only detect up to the fourth-order (four-body) correlations. Generalization to detection of higher order correlations is in principle possible by using more complicated dynamical control (in timing, composition, etc) and/or different types of probes (e.g., higher spins). Exploration along this line will be interesting topics for future studies.

## Method

**Numerical simulation method.** The P-donor electron spin decoherence in a natural abundance $^{29}$Si nuclear spin bath was numerically solved by the well-established cluster-correlation expansion (CCE) method [42]. The central spin coherence time depends on the random configuration of $^{29}$Si nuclear spin positions in the lattice. To compare with the experimental results, we ran simulations for 100 random nuclear spin configurations and took the ensemble average of the corresponding time-domain spin coherence. Since the central spin decoherence is almost independent of the initial state of the nuclear spin bath, we just took a random single-sample state $|J\rangle$ (an eigenstate of $\{I_i^z\}$) as the initial state of the nuclear spin bath.

**Experimental setup.** Experimental results were measured on a natural silicon Czochralski wafer doped with $3\times10^{14}$/cm$^3$ phosphorus, using a BrukerElexsys580 X band (9.6 GHz) spectrometer. All decay times were obtained on the high-field ESR line ($m_I = -1/2$) at 3452 G at 6 K [where the electron spin relaxation processes ($T_1 \approx 1$ s) did not contribute to decoherence over the timescales considered in this paper]. The multiple pulses required for the DD sequences can result in "stimulated echoes", and other unwanted echoes, in the experiment due to pulse infidelities. When such echoes overlap with the desired one (from spin packets which have been flipped by all the π pulses), the experimentally observed decay curves gain unwanted contributions. We therefore cycled the phases of the applied π pulses in such a way as to remove the contribution of all undesired echoes. For UDD, the timings between each pulse are different and most stimulated echoes fall outside the desired one which can then be isolated. For example, the phase cycling sequence for UDD-4 requires simply subtracting the echo from two experiments where the first two pulses are changed from +π to –π and the last two are +π. For CMPG, this is more challenging as the intervals are equal and we did not suppress all possible stimulated echoes for CPMG-5 and CPMG-6.

## References

1. Zurek, W. H. Decoherence, einselection, and the quantum origins of the classical. *Rev. Mod. Phys.* **75**, 715 (2003).
2. Schlosshauer, M. Decoherence, the measurement problem, and interpretations of quantum mechanics. *Rev. Mod. Phys.* **76**, 1267 (2005).
3. Prokof'ev, N. V. & Stamp, P. C. E. Theory of the spin bath. *Rep. Prog. Phys.* **63**, 669 (2000).
4. Awschalom, D. D. *et al*. Quantum spintronics: Engineering and manipulating atom-like spins in semiconductors. *Science* **339**, 1174 (2013).
5. Ladd, T. D. *et al*. Quantum computers. *Nature* **464**, 45 (2010).
6. Hall, L. T. *et al*. Monitoring ion-channel function in real time through quantum decoherence. *Proc. Natl. Acad. Sci. USA* **107**, 18777–18782 (2010).
7. Hall, L. T. *et al*. Ultrasensitive diamond magnetometry using optimal dynamic decoupling. *Phys. Rev. B* **82**, 045208 (2010).
8. Zhao, N., Hu, J. -L., Ho , S. -W., Wan, J. T. K. & Liu, R. B. Atomic-scale magnetometry of distant nuclear spin clusters via nitrogen-vacancy spin indiamond. *Nat. Nanotechnol.* **6**, 242-246 (2011).
9. Zhao, N. *et al.* Sensing single remote nuclear spins. *Nat. Nanotechnol.* **7**, 657-662 (2011).

10. Shi, F. Z. *et al.* Sensing and atomic-scale structure analysis of single nuclear-spin clusters in diamond. *Nat. Phys.* **10**, 21-25 (2014).

11. Feher, G. Electron spin resonance experiments on donors in silicon. I. Electronic structure of donors by the electron nuclear double resonance technique. *Phys. Rev. B* **114**, 1219 (1959).

12. Witzel, W. M., Carroll, M. S., Morello, A., Cywin´ski, Ł. & Das Sarma, S. Electron spin decoherence in isotope-enriched silicon. *Phys. Rev. Lett.* **105**, 187602 (2010).

13. Tyryshkin, A. M., Lyon, S. A., Astashkin, A. V. & Raitsimring, A. M. Electron spin relaxation times of phosphorus donors in silicon. *Phys. Rev. B* **68**, 193207 (2003).

14. de Sousa, R. & Das Sarma, S. Theory of nuclear-induced spectral diffusion: Spin decoherence of phosphorus donors in Si and GaAs quantum dots. *Phys. Rev. B* **68**, 115322 (2003).

15. Witzel, W. M. & Das Sarma, S. Quantum theory for electron spin decoherence induced by nuclear spin dynamics in semiconductor quantum computer architectures: Spectral diffusion of localized electron spins in the nuclear solid-state environment. *Phys. Rev. B* **74**, 035322 (2003).

16. Morton, J. J. L., McCamey, D. R., Eriksson, M. A. & Lyon, S. A. Embracing the quantum limit in silicon computing. *Nature* **464**, 345-353 (2011).

17. Yao, W., Liu, R. -B. & Sham, L. J. Theory of electron spin decoherence by interacting nuclear spins in a quantum dot. *Phys. Rev. B* **74**, 195301 (2006).

18. Liu, R. -B., Yao, W. & Sham, L. J. Control of electron spin decoherence caused by electron–nuclear spin dynamics in a quantum dot. *New J. Phys* **9**, 226 (2007).

19. Gruber, A., Dräbenstedt, A., Tietz, C., Fleury, L., Wrachtrup, J. & von Borczyskowski, C. Scanning confocal optical microscopy and magnetic resonance on single defect centers. *Science* **276**, 2012-2014 (1997).

20. Zhao, N., Ho, S. -W. & Liu, R. -B. Decoherence and dynamical decoupling control of nitrogen vacancy center electron spins in nuclear spin baths. *Phys. Rev. B* **85**, 115303 (2012).

21. Quan, H. T., Song, Z., Liu, X. F., Zanardi, P. & Sun, C. P. Decay of Loschmidt Echo Enhanced by Quantum Criticality. *Phys. Rev. Lett.* **96**, 140604 (2006).

22. Wei, B. -B. & Liu, R. -B. Lee-Yang zeros and critical times in decoherence of a probe spin coupled to a bath. *Phys. Rev. Lett.* **109**, 185701 (2012).

23. Chen, S. -W., Jiang, Z. -F. & Liu, R. -B. Quantum criticality at high temperature revealed by spin echo. *New Journal of Physics* **15**, 043032 (2013).

24. Cai, J., Retzker, A., Jelezko, F. & Plenio, M. B. A large-scale quantum simulator on a diamond surface at room temperature. *Nat. Phys.* **9**, 168-173 (2013).

25. Chemla, D. S. & Shah, J. Many-body and correlation effects in semiconductors. *Nature* **411**, 549-557 (2001).

26. Östreich, Th., Schönhammer, K. & Sham, L. J. Exciton-exciton correlation in the nonlinear optical regime. *Phys. Rev. Lett.* **74**, 4698 (1995).

27. Bolton, S. R., Neukirch, U., Sham, L. J., Chemla, D. S. & Axt, V. M. Demonstration of sixth-order Coulomb correlations in a semiconductor single quantum well. *Phys. Rev. Lett.* **85**, 2002 (2000).

28. Turner, D. B. & Nelson, K. A. Coherent measurements of high-order electronic correlations in quantum wells. *Nature* **466**, 1089-1092 (2010).

29. Ernst, R. R., Bodenhausen, G. & Wokaun, A. *Principles of Nuclear Magnetic Resonance in One and Two Dimensions* (Oxford University Press, USA, 1990).

30. Brixner, T. *et al.* Two-dimensional spectroscopy of electronic couplings in photosynthesis. *Nature* **434**, 625-628 (2005).

31. Li, X., Zhang, T., Borca, C. N. & Cundiff, S. T. Many-body interactions in semiconductors probed by optical two-dimensional fourier transform spectroscopy. *Phys. Rev. Lett.* **96**, 057406 (2006).

32. Bader, S. D. Colloquium: Opportunities in nanomagnetism. *Rev. Mod. Phys.* **78**, 1-15 (2006).

33. Kane, B. E. A silicon-based nuclear spin quantum computer. *Nature* **434**, 625-628 (2005).

34. Viola, L., Knill, E. & Lloyd, S. Dynamical decoupling of open quantum systems. *Phys. Rev. Lett.* **82**, 2417 (1998).

35. de Lange, G., Wang, Z. H., Ristè, D., Dobrovitski,V. V. & Hanson, R. Universal dynamical decoupling of a single solid-state spin from a spin bath. *Science* **330**, 60-63 (2010).

36. Carr, H. Y. & Purcell, E. M. Effects of diffusion on free precession in nuclear magnetic resonance experiments. *Phys. Rev.* **94**, 630 (1954).

37. Meiboom, S. & Gill,D. Modified spin-echo method for measuring nuclear relaxation times. *Rev. Sci. Instrum.* **29**, 688 (1958).

38. Witzel, W. M. & Das Sarma, Multiple-pulse coherence enhancement of solid state spin qubits. *Phys. Rev. Lett.* **98**, 077601 (2007).

39. Uhrig, G. S. Keeping a quantum bit alive by optimized π-pulse sequences. *Phys. Rev. Lett.* **98**, 100504 (2007).

40. Yang, W. & Liu, R. -B. Universality of Uhrig dynamical decoupling for suppressing qubit pure dephasing and relaxation. *Phys. Rev. Lett.* **101**, 180403 (2008).

41. Saikin, S. K., Yao, W. & Sham, L. J. Single-electron spin decoherence by nuclear spin bath- Linked-cluster expansion approach. *Phys. Rev. B* **75**, 125314 (2007).

42. Yang, W. & Liu, R. -B. Quantum many-body theory of qubit decoherence in a finite-size spin bath. *Phys. Rev. B* **78**, 085315 (2008).

**Acknowledgments** The authors thank I. Tchernov for his contributions to the experimental design. This work was supported by Hong Kong RGC/GRF Project 401413, The Chinese University of Hong Kong Focused Investments Scheme, National Basic Research Program of China (973 Program) Grant No. G2009CB929300, and National Natural Science Foundation of China Grant No. 61121491. Work at UCL was supported by the European Research Council under the European Community's Seventh Framework Programme (FP7/2007-2013)/ERC (Grant No. 279781), and by the Engineering and Physical Sciences Research Council (EPSRC) grants EP/K025945/1 and EP/I035536/2. J. J. L. M. is supported by the Royal Society.

**Author Contributions** R.B.L. conceived the idea. W.L.M. and N.Z. performed the theoretical study, G.W. and J.J.L.M. carried out the experimental study. W.L.M., G.W. and R.B.L. wrote the paper. S.S.L. discussed the scheme and the results. All authors analyzed the results and commented on the manuscript.

**Competing financial interests** The authors declare no competing financial interests.

**Correspondence** and requests for materials should be addressed to R.B.L.(email: rbliu@phy.cuhk.edu.hk) or J.J.L.M. (email: jjl.morton@ucl.ac.uk).

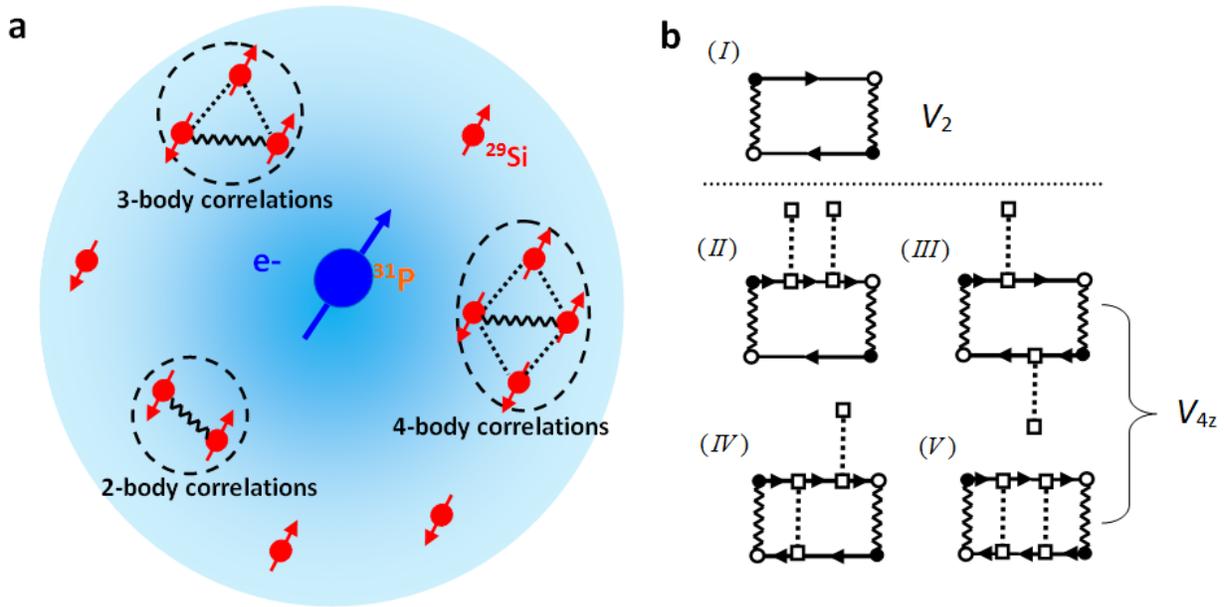

**Figure 1 | Many-body correlations in the $^{29}$Si nuclear spin bath probed by a phosphorus donor electron spin. (a)** Due to the extended donor wavefunction, the P-donor electron spin (blue arrow) interacts with a bath of $^{29}$Si nuclear spins (red arrows) possessing various many-body correlations. **(b)** Topologically inequivalent connected diagrams (LCE diagrams) corresponding to different many-body correlations in the nuclear spin bath: (I) $V_2$-second-order pairwise flip-flop diagram, (II-V) $V_{4z}$-fourth-order diagonal interaction renormalized pairwise flip-flop diagrams. Here the nuclear spin operators $I_i^+$, $I_i^-$, $I_i^z$ are represented in turn by filled circles, empty circles or empty squares. The off-diagonal (diagonal) interaction terms are represented by wavy (dashed) lines. The solid arrows represent nuclear spin correlation functions between $I_i^{\pm}$ and $I_i^{\mp}$ or $I_i^z$ with the arrows indicating the direction of propagation time.

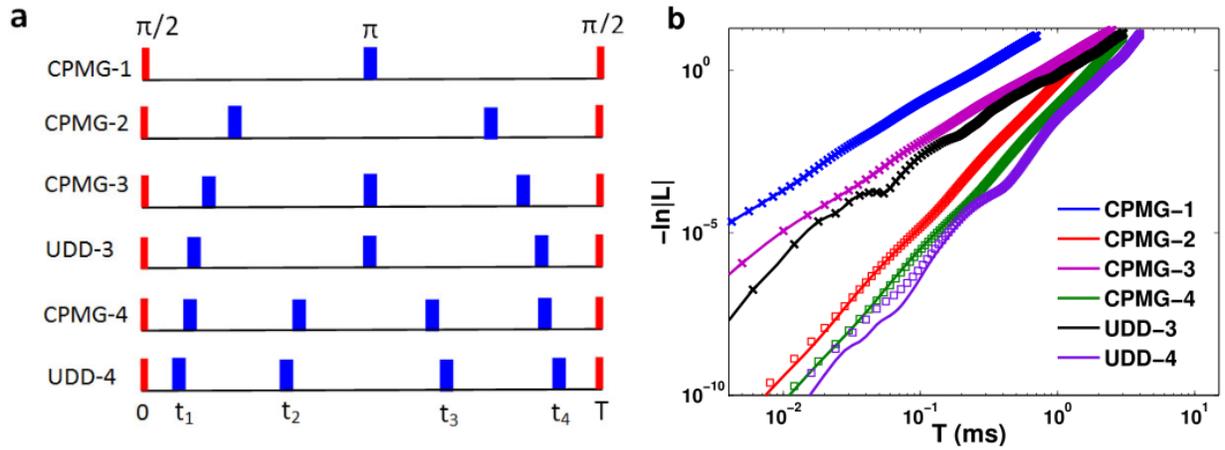

**Figure 2 | Effects of different orders of many-body correlations on central spin decoherence under dynamical decoupling. (a)** Schematics of various CPMG and UDD pulse sequences. **(b)** Comparisons of the P-donor electron spin decoherence in a natural-abundance $^{29}$S inuclear spin bath calculated by the numerically exact CCE method (lines) and those by the LCE approximation (symbols) to determine the many-body correlations that contribute significantly to the spin decoherence under various CPMG and UDD controls. Here, LCE-$V_2$ (crosses) represents the pairwise flip-flop processes in the nuclear spin bath which dominate for sequences with an odd number of π pulses, while LCE-$V_{4z}$ (squares) represents the diagonal interaction renormalized pairwise flip-flop processes which dominate for the even-numbered sequences where LCE-$V_2$ is zero (see Fig. 1b). The magnetic field was set as $B = 0.3$ T applied along the [110] lattice direction.

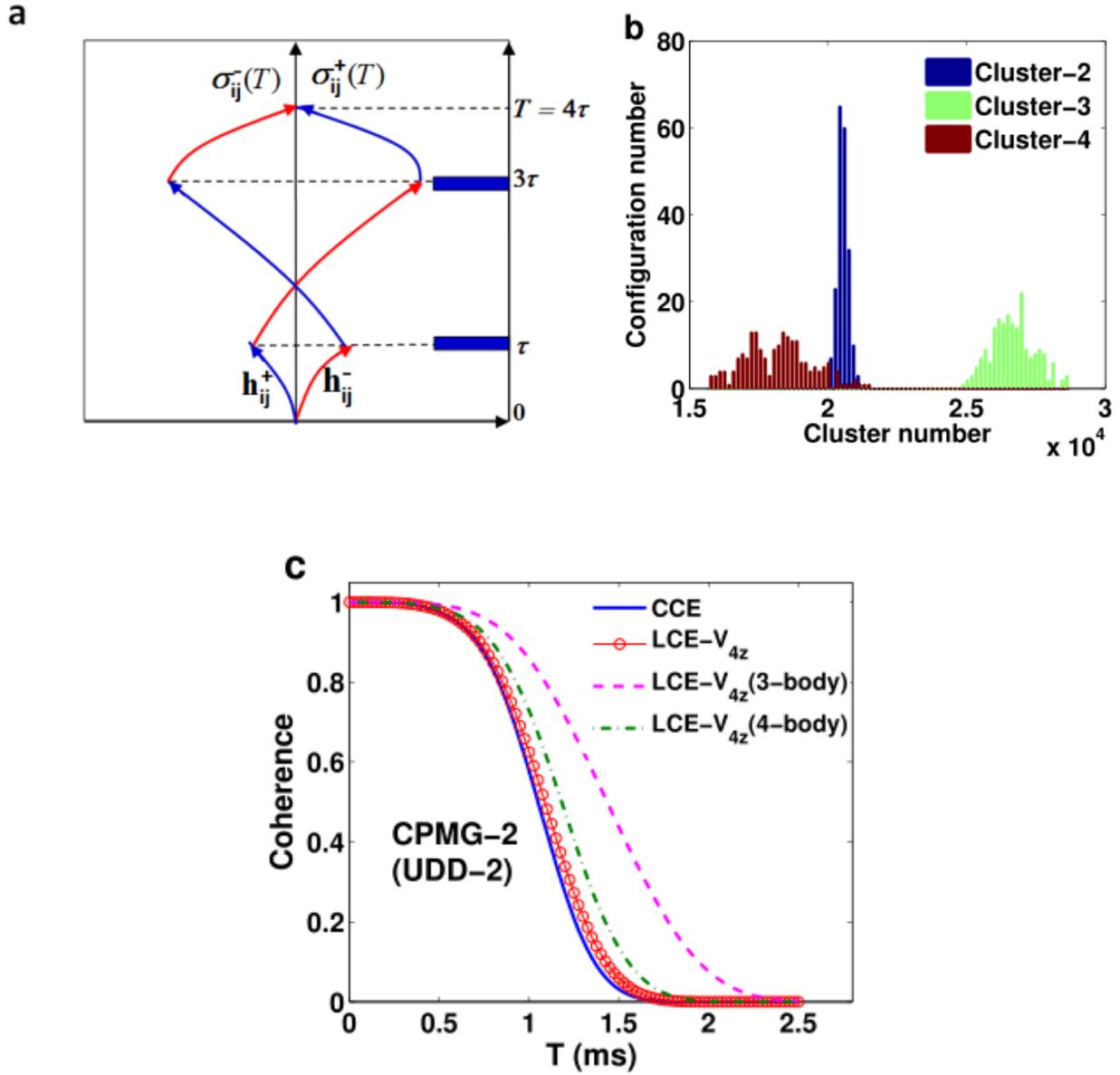

**Figure 3 | Contributions of three-body and four-body correlations to the central spin decoherence under CPMG-2 control. (a)** Schematics of bifurcated pseudo-spin evolutions conditioned on the central spin state under CPMG-2 (or UDD-2) control. The conjugate pseudo-spins $\sigma_{ij}^{\pm}(t)$ (corresponding to the central spin in the state $|\pm\rangle$) describe the dynamics of two-spin correlations. The more the trajectories are separated, the greater the central spin decoherence. The conjugate pseudo-spins exchange their pseudo-fields $h_{ij}^{\pm}$ at time $t = \tau, 3\tau$ when the central

spin is flipped by a π-pulse. Without the diagonal interaction renormalization the conjugate trajectories are symmetric and coincide at time $T$ in the leading order of the evolution time, leading to cancellation of decoherence. **(b)** Histogram of the number of nuclear spin clusters (with inter-nuclei distances < 1 nm) in 200 different bath configurations. **(c)** Decomposition of the LCE-$V_{4z}$ term into three-body and four-body correlations (see Fig 1a) for CPMG-2 (or UDD-2) control of the central spin. The magnetic field was $B = 0.3$ T applied along the [110] lattice direction.

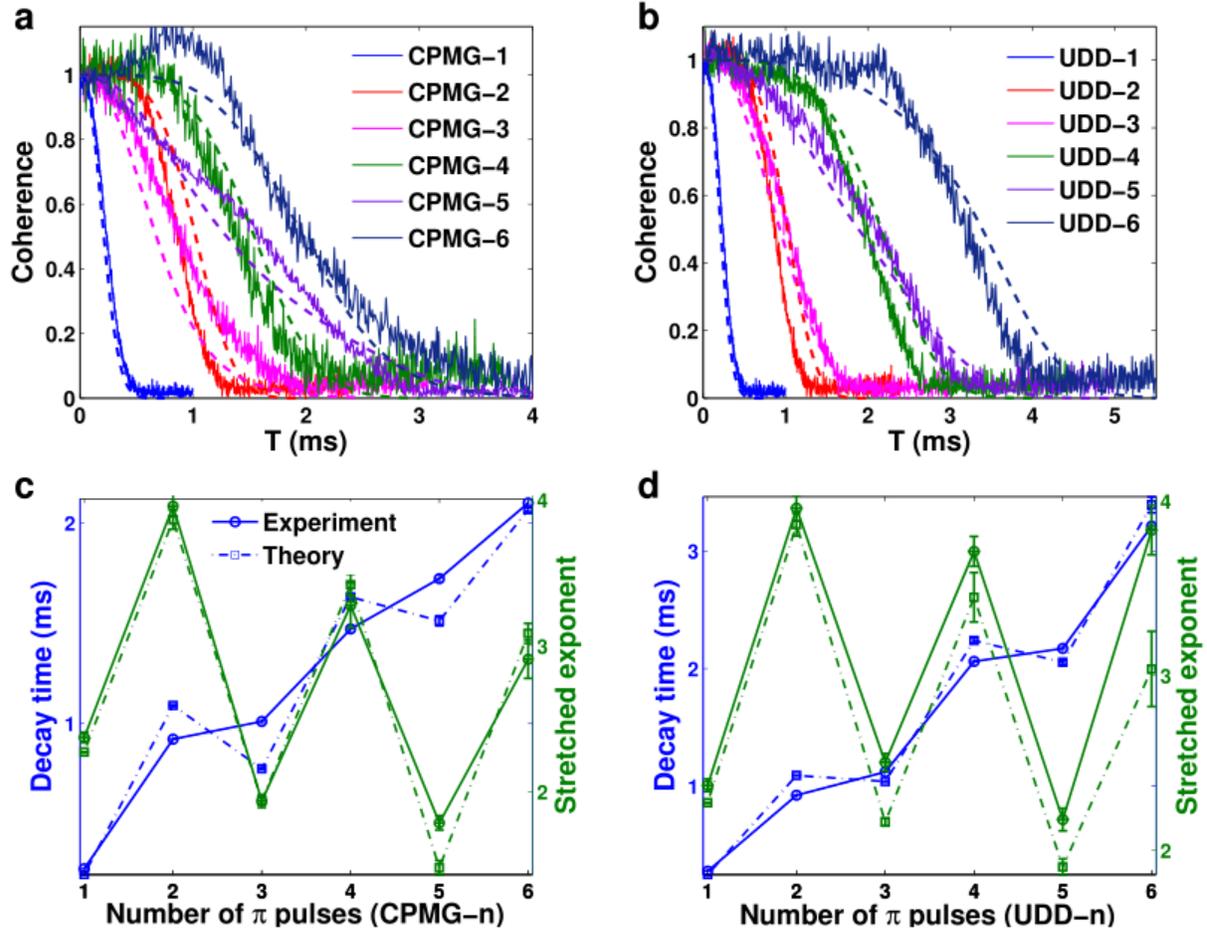

**Figure 4 | Comparison between theoretical and experimental results of $^{nat}$Si:P electron spin decoherence under dynamical decoupling.** (**a,b**) Measured (solid lines) and calculated (dashed lines) coherence of the P-donor electron spin in the natural $^{29}$Si nuclear spin bath under (a) CPMG or (b) UDD control. We attribute the deviation seen at ~1 ms for CPMG-6 to an overlap with uncorrected stimulated/unwanted echoes. (**c,d**) Comparisons of the experimental (solid lines) and theoretical (dashed line) decay times $\tau_{SD}$ (blue) and stretched exponents $\lambda$ (green) of the central spin decoherence under (c) CPMG or (d) UDD control. The magnetic field was $B = 0.3$ T applied along the [110] lattice direction.

# SUPPLEMENTARY INFORMATION

# for

## Uncovering many-body correlations in nanoscale nuclear spin baths by central spin decoherence

Wen-Long Ma, Gary Wolfowicz, Nan Zhao, Shu-Shen Li, John J. L. Morton & Ren-Bao Liu

## I. Analytical Derivation of LCE terms

### A. Interaction picture

The propagators of the nuclear spin bath can be written as [41]

$$\exp[-i(V+H_0)t] = \exp(-iH_0 t)\hat{T}\left\{\exp\left[-i\int_0^t V(t')dt'\right]\right\}$$
$$= \hat{T}\left\{\exp\left[-i\int_0^t V(t'-t)dt'\right]\right\}\exp(-iH_0 t), \quad \text{(S1a)}$$

$$\exp[-i(V-H_0)t] = \exp(iH_0 t)\hat{T}\left\{\exp\left[-i\int_0^t V(-t')dt'\right]\right\}$$
$$= \hat{T}\left\{\exp\left[-i\int_0^t V(-t')dt'\right]\right\}\exp(iH_0 t), \quad \text{(S1b)}$$

where $\hat{T}$ is the time-ordering operator and

$$V(t) = \exp(iH_0 t)V\exp(-iH_0 t) = \sum_{i<j} D_{ij}\left[I_i^+(t)I_j^-(t) + I_i^-(t)I_j^+(t) - 4I_i^z I_j^z\right], \quad \text{(S2)}$$

with $I_i^\pm(t) = I_i^\pm e^{\pm i\omega_i t}$ and $\omega_i = A_i/2$. By the relations above, the operator $U_\pm^n(T)$ can be rewritten in the interaction picture as the product of several evolution operators. For example, for the CPMG-1 (UDD-1) and CPMG-2 (UDD-2) controls

$$U_\pm^1(2t) = \hat{T}\left\{\exp\left[-i\int_0^t b(t'-t)dt'\right]\right\}T\left\{\exp\left[-i\int_0^t b(-t')dt'\right]\right\}, \quad \text{(S3a)}$$

$$U_\pm^2(4t) = \hat{T}\left\{\exp\left[-i\int_0^t b(t-t')dt'\right]\right\} T\left\{\exp\left[-i\int_0^t b(t')dt'\right]\right\} \times \\ \hat{T}\left\{\exp\left[-i\int_0^t b(t'-t)dt'\right]\right\} T\left\{\exp\left[-i\int_0^t b(-t')dt'\right]\right\},$$

(S3b)

with $t = T/(2n)$.

## B. Generalized Wick's theorem for spin 1/2 operators

Wick's theorem for bosons or fermions cannot be directly used for the nuclear spins, because the commutation brackets of spin operators do not yield $c$-numbers. Previous studies generalized Wick's theorem to spin 1/2 operators [41, S1]. First we define the contraction of two spin operators as

$$\overline{I_i^\alpha(t)I_i^\beta(t')} = \hat{T}\{I_i^\alpha(t)I_i^\beta(t')\} - N\{I_i^\alpha(t)I_i^\beta(t')\},$$

(S4)

where $\hat{N}\{\cdots\}$ is the normal-ordered operator depending on the state of the nuclear spin $|\psi\rangle_i$ such that $\hat{N}\{\cdots\}|\psi\rangle_i = 0$. For example,

$$\hat{N}\{I_i^+(t)I_i^-(t')I_i^z(t'')\}|\uparrow\rangle_i = I_i^-(t')I_i^+(t)I_i^z(t'')|\uparrow\rangle_i = 0,$$

(S5a)

$$\hat{N}\{I_i^+(t)I_i^-(t')I_i^z(t'')\}|\downarrow\rangle_i = I_i^+(t)I_i^-(t')I_i^z(t'')|\downarrow\rangle_i = 0.$$

(S5b)

If the nuclear spin $i$ is in the spin-down state ($|\psi\rangle_i = |\downarrow\rangle$), we have the following contraction relations [S1]

$$\overline{I_i^+(t)I_j^-(t';t_1\cdots t_m)} = -2\delta_{ij}I_i^z(t;t')\theta(t'-t)\theta(t_1-t)\cdots\theta(t_m-t)\exp(i\omega_{ij}t), \\ \overline{I_i^-(t;t_1\cdots t_m)I_j^z(t';t'')} = -\delta_{ij}I_i^-(t;t_1\cdots t_m t')\theta(t''-t), \\ \overline{I_i^+(t)I_j^z(t';t'')} = \delta_{ij}I_i^+(t)\theta(t'-t).$$

(S6)

where $\theta(t)$ is the Heaviside step function. If $|\psi\rangle_i = |\uparrow\rangle$, we can get the new contraction relations from (S6) by the transformation $I_i^\pm \to -I_i^\mp$.

Now we can state the generalized Wick's theorem for spin 1/2 operators: the time-ordered product of a set of time-dependent spin operators is equal to the sum of all possible fully contracted products which contains only $I_i^z$ operators [41, S1].

## C. Derivation of LCE terms

Now we can derive the analytical forms of the LCE terms. First we calculate the LCE-$V_1$ term [see Fig. S1(a)],

$$\langle V_1 \rangle = \int_C \langle J | \hat{T}_C \{V(t_1)\} | J \rangle dt_1 = \sum_{ij}(-4D_{ij}) \int_C \langle ij | I_i^z I_j^z | ij \rangle dt_1 = 0. \tag{S7}$$

where $|J\rangle = \otimes |j\rangle$ and $|ij\rangle = |i\rangle \otimes |j\rangle$. We see that this term vanishes due to the contour integral.

The LCE-$V_2$ term [see Fig. S1(b)] is

$$\begin{aligned}\langle V_2 \rangle &= \frac{1}{2!}\int_C dt_1 dt_2 \langle J | \hat{T}_C \{V(t_1)V(t_2)\} | J \rangle \\ &= \sum_{i>j}\int_C dt_1 dt_2 \langle ij | \hat{T}_C \{I_i^+(t_1)I_j^-(t_1)I_j^+(t_2)I_i^-(t_2)\} | ij \rangle \\ &= \sum_{i>j}\int_C dt_1 dt_2 \langle ij | \overbrace{I_i^+(t_1)I_j^-(t_1)I_j^+(t_2)I_i^-(t_2)} | ij \rangle .\end{aligned} \tag{S8}$$

For the CPMG-$n$ control, we have $\mathrm{Re}\langle V_2 \rangle = \sum_{ij} 4D_{ij}^2 \omega_{ij}^{-2}\left[4\cos(\omega_{ij}t) - \cos(2\omega_{ij}t) - 3\right]$ when $n$ is odd, and $\mathrm{Re}\langle V_2 \rangle = 0$ when $n$ is even. For the UDD-$n$ control, we also have $\mathrm{Re}\langle V_2 \rangle = 0$ when $n$ is even, but $\mathrm{Re}\langle V_2 \rangle$ cannot be written in a simple compactform as in the CPMG case when $n$ is odd ($n > 2$).

The LCE-$V_{4z}$ term includes four diagrams [Fig. S1(g-j)]. However, the last two diagrams [Fig. S1(i-j)] have little contribution to central spin decoherence, because the pairwise flip-flop processes of nuclear spins $(i, j)$ are independent of the diagonal interactions between them

($D_{ij}I_i^z I_j^z$) [so the 4-th order terms in Fig. S1(i)-(j) approximately reduce to the same form as in Fig. S1(c)-(d), respectively, but are higher-order small quantities]. For the diagrams in Fig. S1(g-h), we can get analytical results of the three-body and four-body correlations for the CPMG and UDD control of even pulse number as follows

$$-\ln\left|L_{ijk}\right| \sim \left\langle I_k^z\right\rangle^2 \frac{D_{ij}^2}{\omega_{ij}^4}\left(D_{ik} - D_{jk}\right)^2,$$

$$-\ln\left|L_{ijkl}\right| \sim \left\langle I_k^z\right\rangle\left\langle I_l^z\right\rangle \frac{D_{ij}^2}{\omega_{ij}^4}\left(D_{ik} - D_{jk}\right)\left(D_{il} - D_{jl}\right),$$

(S9)

where $L_{ijk}$ and $L_{ijkl}$ denote the central spin decoherence caused by the diagonal interaction renormalized pairwise flip-flop processes ($i \leftrightarrow j$) in the three-spin cluster $\{i,j,k\}$ [Fig. S2(b)] and four-spin clusters $\{i,j,k,l\}$ [Fig. S2(c)], respectively, and $\left\langle I_k^z\right\rangle \equiv \left\langle J|I_k^z|J\right\rangle$. These analytical expressions imply that to have significant contributions to the central spin decoherence the nuclear spin clusters should satisfy the following conditions: (i) the inter-nuclei distances in four-spin clusters or three-spin clusters should be rather small (<1 nm); (ii) the renormalization to the energy cost of the pair flip-flop ($i, j$) should be substantial as compared with the bare energy cost, i.e., $\left|\omega_{ij}^{-1}\left\langle I_k^z\right\rangle\left(D_{ik} - D_{jk}\right)\right|$ should be large for three-spin clusters $\{i,j,k\}$ while $\omega_{ij}^{-2}\left\langle I_k^z\right\rangle\left\langle I_l^z\right\rangle\left(D_{ik} - D_{jk}\right)\left(D_{il} - D_{jl}\right)$ should be positive and large for four spin clusters $\{i,j,k,l\}$.

## II. Pseudo-spin Model

To get an intuitive understanding of the pulse-number parity effect, we use the pseudo-spin model [17] to describe the dynamics of two nuclear spins. In the strong field regime, the Hamiltonian of the *i*-th and *j*-th nuclear spins conditioned on the central spin state

$$H_\pm^{ij} = \pm\omega_{ij}\sigma_z/2 + D_{ij}\sigma_x,$$

(S10)

where the basis set is defined as $\{|\uparrow\downarrow\rangle, |\downarrow\uparrow\rangle\}$. Note that the two pseudo-fields corresponding to the two opposite central spin states lie in the *xz*-plane and are symmetric with respect to the *x*-axis. The time evolution operator is

$$U_{\pm}(t) = \cos\phi - i(n_x \sigma_x \pm n_z \sigma_z)\sin\phi, \tag{S11}$$

where $\phi = \kappa t$, $\kappa = \sqrt{\omega_{ij}^2/4 + D_{ij}^2}$, $n_x = D_{ij}/\kappa$, $n_z = \omega_{ij}/\kappa$. If the central spin is under CPMG-*n* or UDD-*n* control, the time evolution operator $U_{\pm}^n$ can be obtained by the above formula. For CPMG-1 (UDD-1) and CPMG-2 (UDD-2) controls, we have

$$\begin{aligned}U_{\pm}^1(T) &= 1 - 2n_x^2 \sin^2\phi - in_x(2n_z \sin^2\phi \sigma_y \mp \sin 2\phi \sigma_x), \\ U_{\pm}^2(T) &= 1 - 2n_x^2 \sin^2 2\phi - 2in_x \sin 2\phi \left[(1 - 2n_x^2 \sin^2\phi)\sigma_x \mp 2n_x \sin^2\phi \sigma_z\right].\end{aligned} \tag{S12}$$

For the donor spin in silicon, we have $\omega_{ij} \gg D_{ij}$, so $n_x \approx 2D_{ij}/\omega_{ij}$ is a small quantity. The difference between $U_{+}^n(T)$ and $U_{-}^n(T)$ causes the central spin decoherence $L_{+,-}^n(T)$. When $n = 2k+1$, we have $|U_{+}^{2k+1} - U_{-}^{2k+1}| \sim n_x$ and $L_{+,-}^{2k+1}(T) \approx 1 - n_x^2 f_{2k+1}(T)$. However, when $n = 2k$, due to the symmetry between the two pseudo-fields corresponding to the two opposite central spin states, the two conjugate trajectories of the pseudo-spin under the two pseudo-fields cross into each other (in the leading order of evolution time) at the end of the DD control. Therefore $|U_{+}^{2k} - U_{-}^{2k}| \sim n_x^2$ and $L_{+,-}^{2k}(T) \approx 1 - n_x^4 f_{2k}(T)$. Here $f_n(T)$ is a function of the total evolution time $T$ and the pulse number of DD control $n$.

If we consider all the nuclear spins in the bath, then the central spin decoherence can be expressed as the product of the decoherence contributed by each pair of nuclear spins. Then we have

$$L_{+,-}^{2k+1}(T) \approx \prod_{ij}\left(1 - \frac{4D_{ij}^2}{\omega_{ij}^2} f_{2k+1}^{ij}(T)\right) \approx \prod_{ij} \exp\left(-\frac{4D_{ij}^2}{\omega_{ij}^2} f_{2k+1}^{ij}(T)\right), \tag{S13a}$$

$$L_{+,-}^{2k}(T) \approx \prod_{ij}\left(1 - \frac{16D_{ij}^4}{\omega_{ij}^4} f_{2k+1}^{ij}(T)\right) \approx \prod_{ij} \exp\left(-\frac{16D_{ij}^4}{\omega_{ij}^4} f_{2k+1}^{ij}(T)\right). \tag{S13b}$$

These results are consistent with results obtained by the LCE method. Recall that the LCE-$V_l$ terms are proportional to $(D_{ij}/\omega_{ij})^l$. Therefore, for CPMG or UDD control of odd pulse numbers, the second-order correlations contribute the most to the central spin decoherence. But for the CPMG or UDD control of even pulse numbers, the second-order correlations are cancelled and the fourth-order correlations corresponding to the ring diagrams $V_{4r}$ and locked diagrams $V_{4l}$ (see Fig S1) would contribute the most to the central spin decoherence. It should be pointed out that in the discussion above we have not considered the diagonal interactions between the nuclear spins $i, j$ and other nuclear spins in this pseudo-spin model. Actually such diagonal interactions will renormalize the pseudo-spin Hamiltonian and break the symmetry between the two conjugate pseudo-fields for the pseudo-spin. Therefore, the diagonal interaction renormalized pairwise flip-flop (instead of $V_{4l}$ and $V_{4r}$) would be the dominant contribution to the central spin decoherence when the number of pulses is even.

## Supplementary References

S1. Giovannini, B. & Koide, S. Perturbation theory for magnetic impurities in metals. *Prog. Theor. Phys.* **34**, 705-725 (1965).

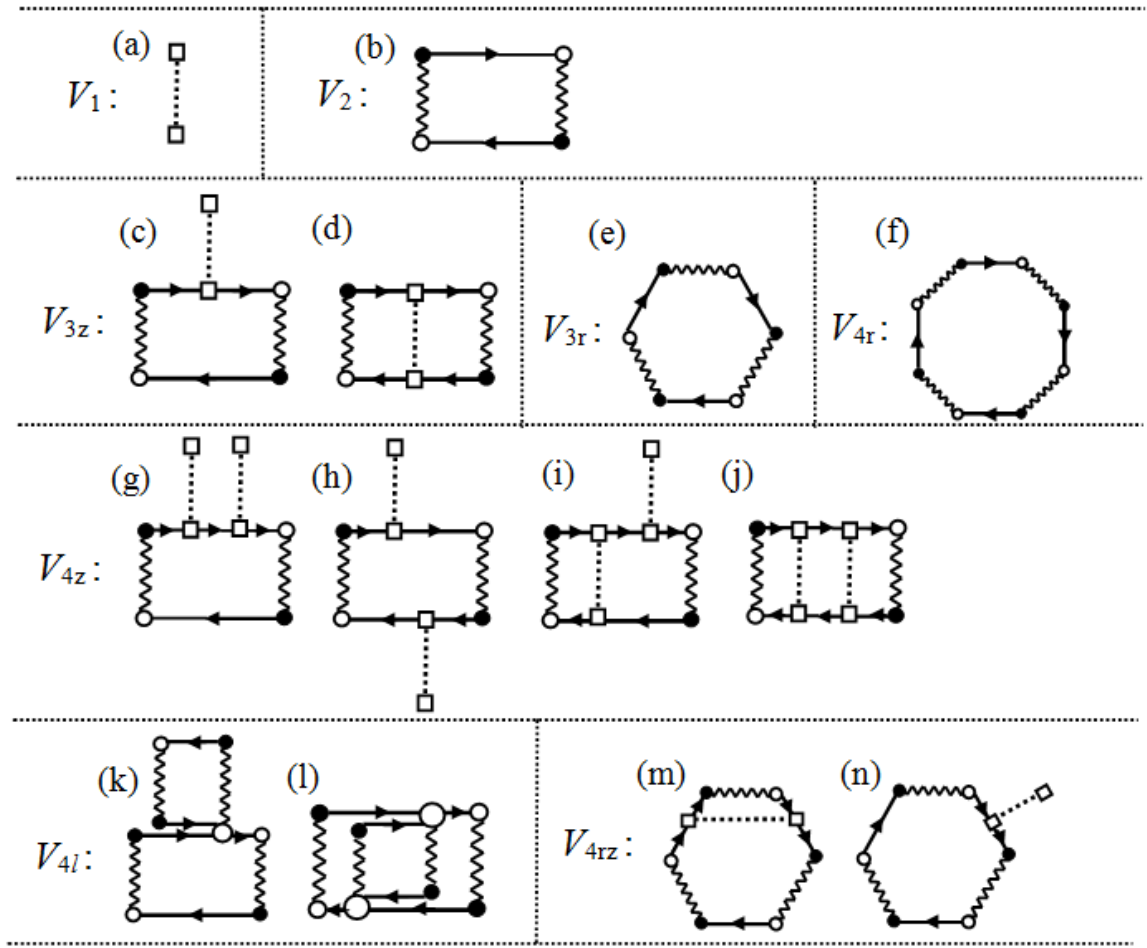

**Figure S1| Topologically inequivalent connected diagrams corresponding to different many-body correlations in the nuclear spin bath up to the fourth order**. (a) $V_1$-first-order diagram, (b) $V_2$-second-order pairwise flip-flop diagram, (c)-(d) $V_{3z}$-third-order diagonal interaction renormalized pairwise flip-flop diagrams, (e) $V_{3r}$-third-order ring diagram, (f) $V_{4r}$-fourth-order ring diagram, (g)-(j) $V_{4z}$-fourth-order diagonal interaction renormalized pairwise flip-flop diagrams, (k)-(l) $V_{4l}$-fourth-order locked diagrams. (m)-(n) $V_{4rz}$-fourth-order diagonal interaction renormalized ring diagrams.

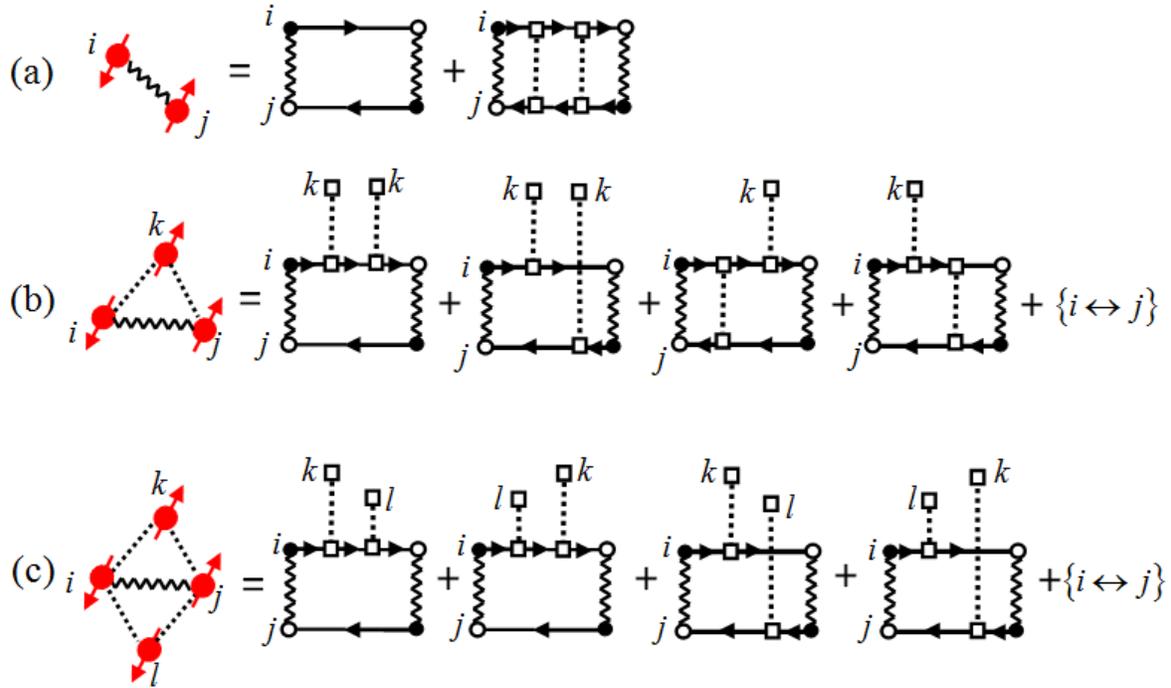

**Figure S2| Decomposition of many-body correlations into LCE diagrams.** We only consider the $V_2$ and $V_{4z}$ terms contributing most to central spin decoherence. The fourth-order diagonal-interaction renormalized pair flip-flop processes ($V_{4z}$) can be two-body, three-body or four-body correlations. The two-body correlations describe the pairwise flip-flop processes of nuclear spins $i, j$ renormalized by the diagonal couplings between $i$ and $j$ while the three-body (four-body) correlations describe the pairwise flip-flop processes of nuclear spins $i, j$ renormalized by the diagonal couplings of $i, j$ to nuclear spin $k$ ($k, l$) in the nuclear spin bath. Note that in this figure the vertices along the same horizontal line are of the same spin.

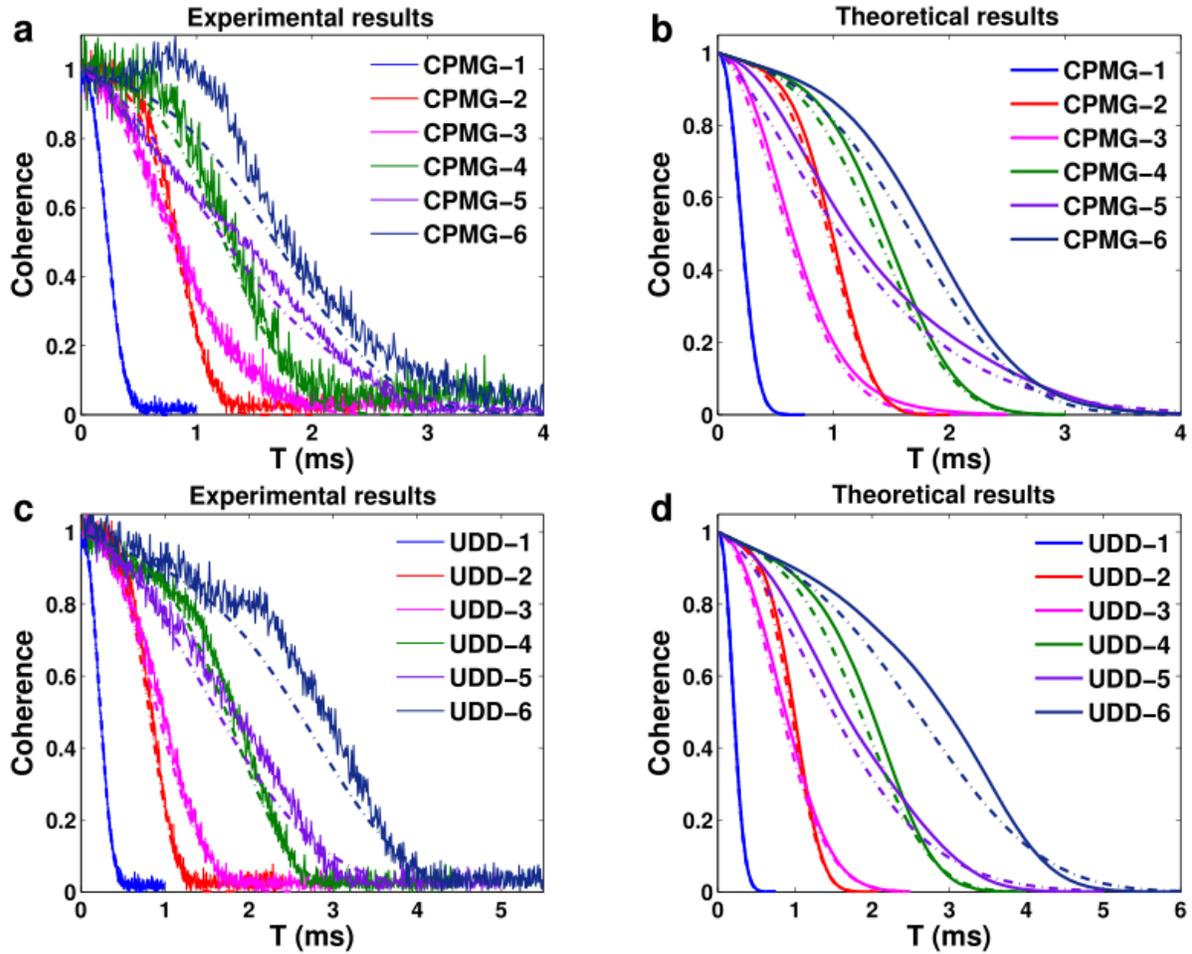

**Figure S3 | Numerical fits of experimental and theoretical results of $^{nat}$Si:P electron spin decoherence by exponential functions $\exp\left[-T_r/T_{ID}/(\tau_{SD})^\lambda\right]$.** (a,c) Experimental or (b,d) theoretical (solid lines) and fitted (dashed lines) coherence of the P-donor electron spin in the natural $^{29}$Si nuclear spin bath under (a,b) CPMG or (c,d) UDD control. Here the same value of $\tau_{ID} = 10$ ms was used in all the fits. We attribute the deviation seen at ~1 ms for CPMG-6 to an overlap with uncorrected stimulated/unwanted echoes. The magnetic field was $B = 0.3$ T applied along the [110] lattice direction.